# Local fluctuations in an aging glass

K. S. Sinnathamby, H. Oukris, and N. E. Israeloff

Dept. of Physics, Northeastern University, Boston, MA 02115

Polarization fluctuations were measured in nanoscale volumes of a polymer glass during aging following a temperature quench through the glass transition. Statistical properties of the noise were studied in equilibrium and during aging. The noise spectral density had a larger temporal variance during aging, i.e. the noise was more non-Gaussian, demonstrating stronger correlations during aging.

75.50.Lk, 75.10.Nr, 77.22.-d, 77.22.Gm, 77.84.Jd, 78.30.Ly

Recent work has deepened our understanding of glass transitions and glassy dynamics[1,2]. Yet a comprehensive theory remains elusive. Meanwhile, remarkable new glassy phenomena are being discovered. One of the most intriguing behaviors in glasses has been aging, which is characterized by slow relaxation of measured properties such as mechanical, magnetic, or dielectric response, following a temperature quench[3]. Scaling of the response curve vs. time or frequency with the age of the system relative to the quench, continues to be of interest[4]. Recently striking memory and rejuvenation effects have been found[3,5]. And the slow return to equilibrium during aging, has made glass models ideal for studying nonequilibrium statistical mechanics, especially the development of a nonequilibrium fluctuation-dissipation-relation (FDR)[6].

An equally important recent development as been the discovery that dynamics in structural glasses are spatially heterogeneous[2, 7, 8], with transient heterogeneities 1 to 3 nm in size[9]. In molecular dynamics simulations[10] and experiments on colloidal glasses[11], fast and slow regions were directly observed. More recently, local probes[8, 12] have been useful for following the dynamics of individual heterogeneities or single probe molecules.

In a confluence of these recent developments, a recent PRL explored *local* aging dynamics, FDR violations, and heterogeneity in a Monte-Carlo simulation of an Edwards-Anderson Ising spin glass model[13]. Local fluctuations, responses, and their ratio, which determines an effective temperature, $T_{eff}$, were found to be spatially heterogeneous during aging. For the young regime, i.e. when the observation time $t-t_w$ was comparable to or longer than the waiting time following the quench, $t_w$, the $T_{eff} > T$, consistent with previous work. Most striking was that when sampled spatially the local $T_{eff}$ were highly heterogeneous, mainly due to more broadly distributed fluctuations as compared to the narrowly distributed responses, or the near equilibrium fluctuations[14]. It was later proposed that the broad distribution of non-equilibrium fluctuations may be characterized by intermittency, e.g. rare but large fluctuation events, arising from the infrequent traversal of large barriers during aging. A recent macroscopic experiment confirms that aging is heterogeneous [15]. In this paper we have used a local probe of dielectric response to measure non-equlibrium fluctuations in a mesoscopic polymer glass during aging just below the glass transition. We find that the local fluctuations during aging are non-Gaussian. The noise measured as a function of time exhibits large modulations, which however, decrease dramatically with waiting time. This indicates more correlated dynamics during aging.

The samples studied were 0.5 -1.0 µm thick poly-vinyl-acetate (PVAc) films with glass transition: $T_g = 308$ K. The films were prepared by spin-coating from solution onto a substrate and dried[16, 17] at 340 K for 20 minutes. The macroscopic aging behavior was studied by sandwiching the PVAc films between sputtered thin AuPd crossed electrodes to form a capacitor. Using a Novocontrol based dielectometer the dielectric susceptibility was measured as a function of frequency, f, and waiting time, $t_w$ following a rapid temperature quench. See figure 1. Aging continues to much larger values of $ft_w$ as compared with spin glasses. For example at 302 K, measuring the complex susceptibility at 1 Hz, the aging continues well beyond 20 hours.

By sensing local electrostatic forces, non-contact atomic force microscopy (AFM) techniques can be used to measure local dielectric properties[16]. In the present experiment, we employed a noise measurement scheme[8, 17] (see Fig. 1 inset), which utilized a small cantilever with a sharp (40 nm radius) conductive tip, mounted in a temperature controlled can within a vacuum chamber. The cantilever is driven at its resonance frequency (~200 kHz), via a piezoelectric slab and a self-driven oscillator circuit. When a voltage bias is applied between tip and substrate, fluctuations in the sample polarization beneath the tip produce proportional variations in the cantilever resonance frequency which are detected using frequency modulation techniques. For small tip heights, z, above the sample surface, the resonance frequency is most sensitive to a small region directly beneath the tip. For z =30 nm and bias = 4V, the effective probed volume: $\Omega \sim 2\times10^{-17}$ cm$^3$, and depth: 40 nm. This volume contains of order $10^3$ heterogeneities, 3 nm in diameter.

With tip height fixed using feedback to reset the resonance frequency between each set of data points (65 s), time-series of PVAc polarization fluctuations were recorded at various temperatures just below $T_g$, and as a function of waiting time, $t_w$, following a temperature quench from 315 K. From each time series begun at waiting time $t_w$ the power spectra $S(f, t_w)$, for f = 0.015 Hz to 15 Hz, were obtained. The decrease of the noise power with aging is comparable to that of the susceptibility as shown in figure 1. To study fluctuations in the power spectra, the local spectral exponent, $\alpha = -d\ln(S)/d\ln(f)$ was measured over typically a decade in frequency in averages of 4 spectra. Then the variations in $\alpha$ were measured as a function of waiting time $t_w$, and analyzed in detail[8]. This technique measures changes in the shape of the noise spectrum with time and is less sensitive to extrinsic sources of variation in the noise magnitude such as drifts in tip height or feedback effects. Simulated Gaussian noise spectra, and spectra from a Monte-Carlo simulation (discussed below) were analyzed in the same way for comparison.

In figure 2a we show $\alpha$ vs. $t_w$ at 302 K. Each $\alpha$ was measured over 0.1 Hz to 1 Hz in a spectrum averaged for 4.3 minutes. Large variations in $\alpha$ occur, particularly at the earlier $t_w$. The variance was typically double the Gaussian expectation. The temporal variance of $\alpha$, $\sigma^2(\alpha)$, was measured for 0.1 Hz to 1 Hz over intervals of $\Delta t_w$ = 2.3 hours, and was averaged over 12 quenches (Figure 2b). Note that the small systematic change of $\alpha$ with aging makes a negligible contribution to the variance. The non-Gaussian portion of the variance, $\sigma_N^2$, obtained by subtracting the Gaussian variance, is plotted (fig. 4) vs. $t_w$ for several temperatures. The variance clearly decreases with waiting time for all temperatures. The decay of variance occurs on $t_w$ scales similar to the decay of susceptibility. The non-Gaussian variance appears to reach a steady value, still in excess of Gaussian, in equilibrium, which decreases with increasing temperature.

How do these non-Gaussian variations in the noise arise? First, we consider a simple heterogeneous model in which the noise in equilibrium is produced by a superposition of Lorentzians, corresponding to Debye peaks in the dielectric susceptibility, whose characteristic frequencies are randomized on a time scale corresponding to the heterogeneity lifetime, $\tau_{het}$. A Monte Carlo simulation of this model gives fluctuations in $\alpha$, whose variance is inversely proportional to the mean density of Lorentzians per unit logarithm of bandwidth, $n_e$. For 1/f noise this number is independent of frequency. In fig. 3, the equilibrium $\sigma_N^2$ clearly decreases with increasing temperature. When compared with the simulation, this corresponds to an $n_e$ which grows with temperature. This is consistent with the fact that the $\alpha$–peak in the distribution of rates moves to higher frequencies with increasing temperature, increasing the number of heterogeneities active in the measurement band.

We can try to extend this simple heterogeneous model, together with the notion of fictive temperature[18], to the aging regime. Initially after a temperature quench, the distribution of rates, as measured in $\varepsilon''(\omega)$, is similar to that of a higher temperature, i.e. a fictive temperature. Thus we might expect the $\sigma_N^2(t_w)$ to be initially smaller, then *rise* toward the equlibrium value. Instead the opposite occurs. The variations during aging are then not consistent with a smoothly decreasing Lorentzian density. It would seem that the noise during aging is determined by fewer independent noise sources. How can this occur? As found in the spin glass simulations [13], the effective temperatures during early aging were very heterogeneous. The regions with higher effective temperatures would be noisier, perhaps dominating the spectrum and reducing the effective number of noise sources. However, in a simple activated two-level-system model with a broad distribution of energies, heterogeneous effective temperatures always *increase* the effective number of noise sources. Alternatively, the usual picture of aging involves traversal of large barriers in the energy landscape. These rare traversals could produce correlated bursts of noise, or freeze the activity in several nearby cooperative regions. In other words the noise during aging may be more correlated, with correlations decreasing with waiting time.

Within the context of the simplest heterogeneous model, dynamics are correlated over the CRR length scale with characteristic size $\xi$. The Lorentzian density may be expected to scale[8] as $n_e = \varepsilon''\Omega/\varepsilon'(0)\xi^3$, while the excess variance, $\sigma_N^2$, scales as $n_e^{-1}$. Since $\varepsilon'(0)$ changes very weakly with temperature and waiting time, we can deduce how an effective dynamical correlation length scales with temperature and waiting time:

$$\xi \approx [\sigma_N^2 \varepsilon'']^{1/3} \qquad (1)$$

In equilibrium (fig. 4 inset), the correlation length decreases weakly with increasing temperature, consistent with some models[19]. During aging, as shown in figure 4, the correlation length decreases similarly with waiting time for all temperatures, and is 50 % larger than the equilibrium value at early $t_w$. This strongly suggests that aging in structural glasses proceeds by events which occur over larger length scales than in equilibrium. In an aging colloidal glass experiment, the CRR length scale was found to remain roughly constant during aging, though it was expected to increase along with relaxation time[20].

In summary, we find a new mesoscopic scale signature of aging in a structural glass. Specifically, fluctuations exhibit a larger temporal variance in a mesoscopic sample during aging than in equilibrium. This can be understood, most simply, as stronger correlation between fluctuations in neighboring regions, i.e. a longer dynamical correlation length. Whether this can be explained in terms of a broadened distribution of sample ages, or effective temperatures is an open question.

This work was supported by NSF DMR-0205258 and ACS PRF 38113-AC7.

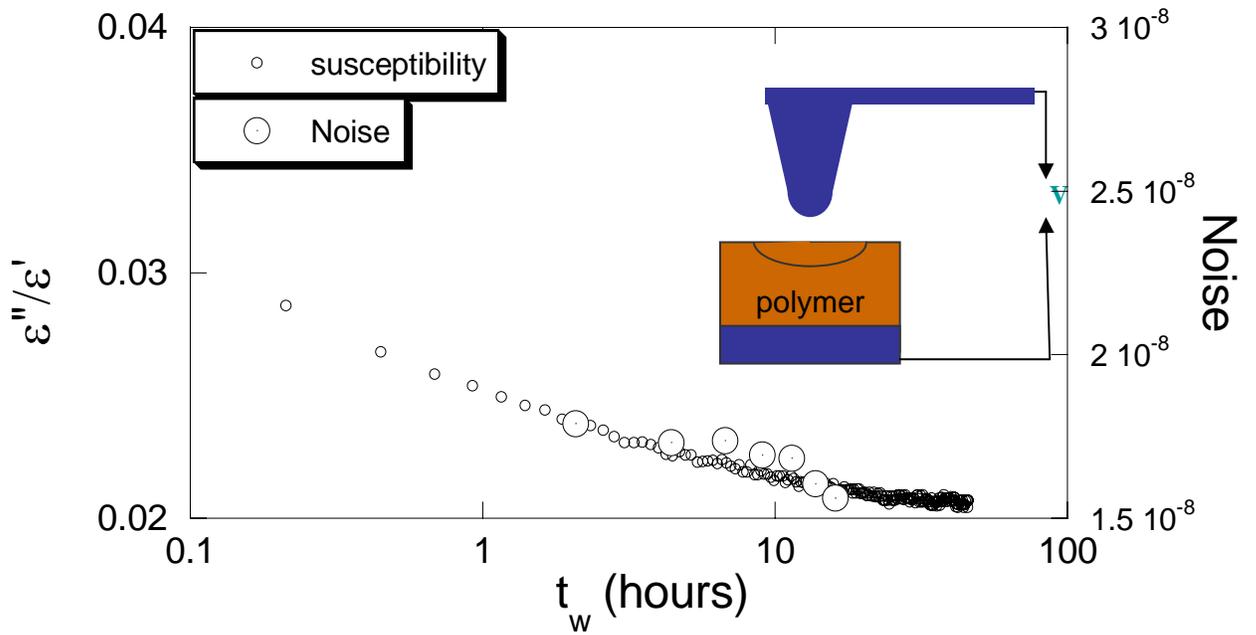

**Figure 1.** The dielectric susceptibility (ratio of imaginary to real components) measured at 1 Hz in a macroscopic sample is plotted vs. waiting time following a temperature quench from 315 K to 302 K. The polarization noise measured during aging in a nanoscopic sample in a band centered on 1.4 Hz at 302 K is also shown. Inset: noise measurement setup showing SPM tip, sample and probed volume.

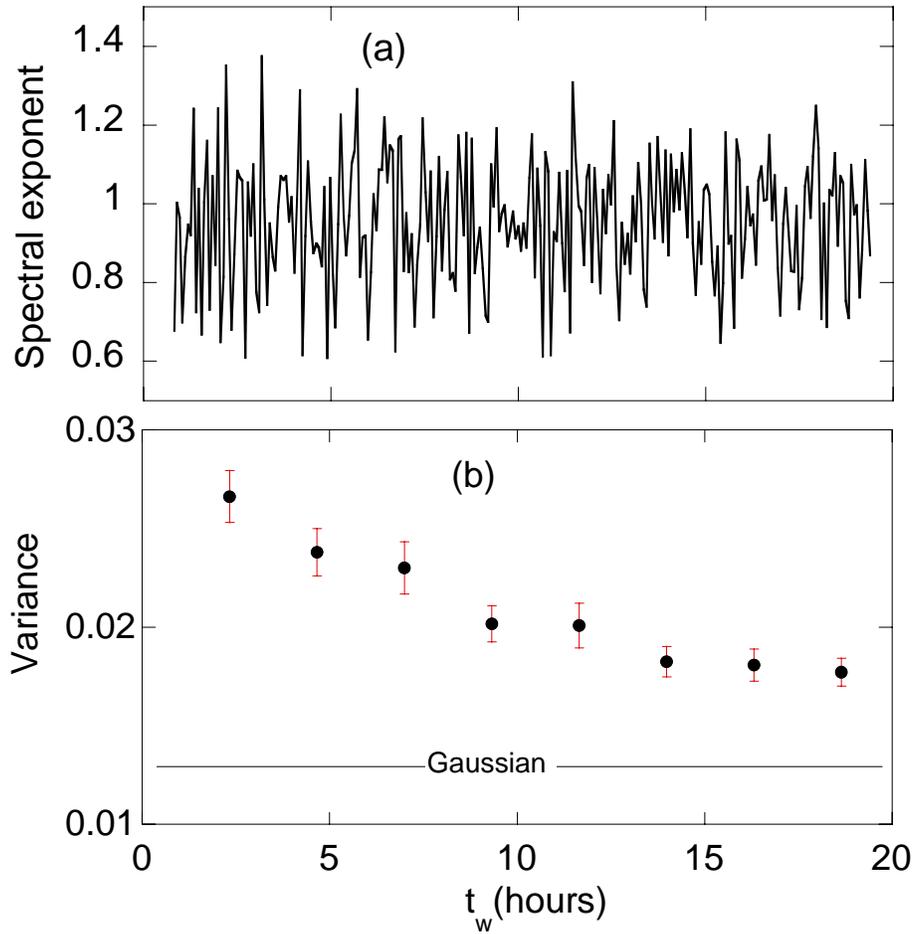

Figure 2. The spectral exponent, measured by a power-law fit between 0.1 Hz and 1 Hz, is plotted (a) vs. waiting time following a quench to 302 K. In (b) the variance of spectral exponent measured over 2.3 hour windows and averaged over 12 quenches is plotted vs. waiting time for 302 K. The prediction for Gaussian noise is shown.

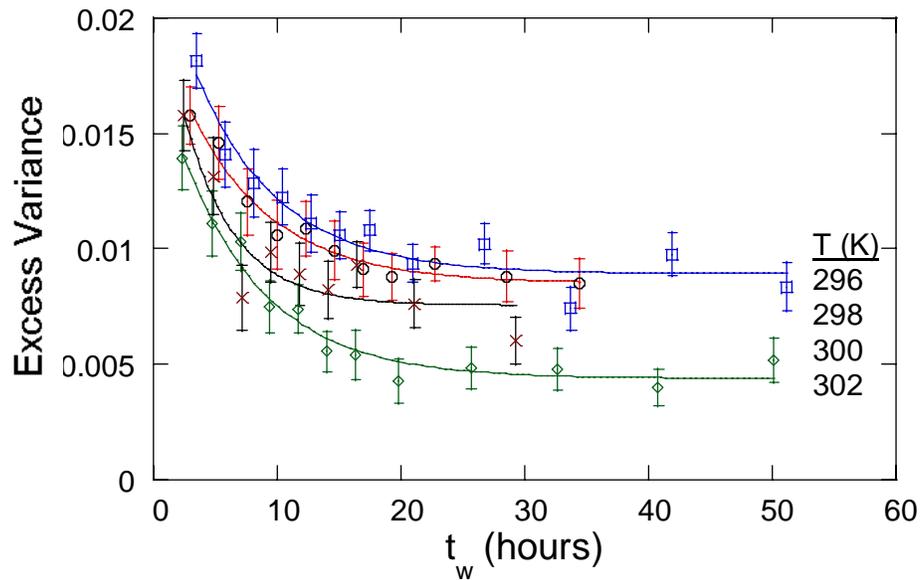

**Figure 3.**

**The excess (above Gaussian) variance in spectral exponent (0.1 Hz-1 Hz) is plottted vs. waiting time for several temperatures (indicated). For waiting times < 20 h the data are averaged over 2.3 h windows and 10 to 20 quenches. After 20 h the variances from several adjacent 2.3 h windows are averaged for 3 to 5 quenches.**

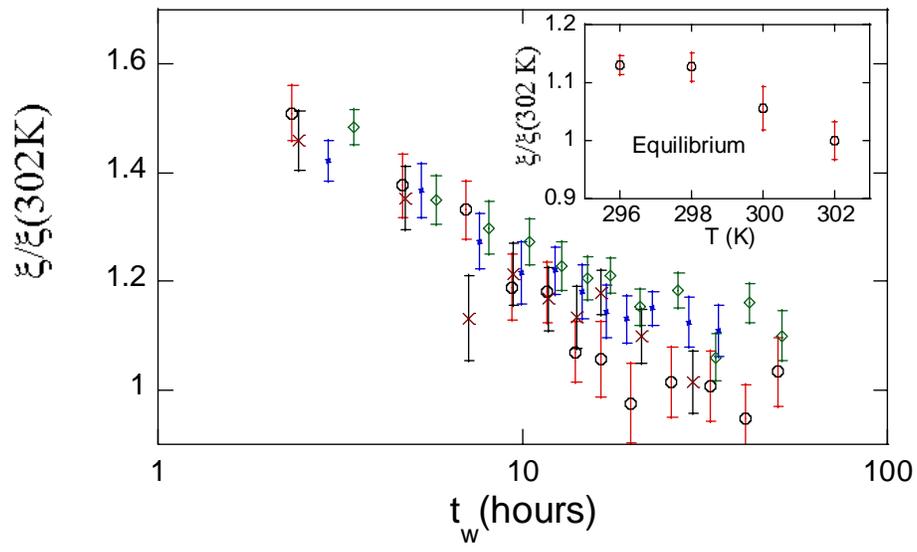

**Figure 4.**

**Nonequilibrium dynamical correlation length vs. waiting time, derived from equation 1, and normalized to the equilibrium value at 302 K. Inset: the equilibrium correlation length grows weakly with decreasing temperature.**